\documentclass[a4paper,oneside,10pt]{article} 
\usepackage{authblk}

\usepackage{cite}
\usepackage{graphicx}  
\usepackage{dcolumn}   
\usepackage{bm}        
\usepackage{amssymb}   
\usepackage{amsmath}
\usepackage{epsfig}
\usepackage{xspace}
\usepackage{url}
\usepackage{hyperref}
\usepackage{color}
\usepackage{ulem}
\usepackage[utf8]{inputenc}

\newcommand{\bea}{\begin{eqnarray}}
\newcommand{\eea}{\end{eqnarray}}

\begin{document}

\title{Accurate Sampling from Diffusion Models}

\author[1]{D\'enes Sexty}

\affil[1] {\it Institute of Physics, NAWI Graz, University of Graz, Universit\"atsplatz 5, 8010 Graz, Austria }


\maketitle

\date{}

\begin{abstract} 
A new proposal called DM-SMC (Diffusion Model - Sequential Monte Carlo) is investigated, which samples ensembles defined in terms of an action, using diffusion models trained on samples from the ensemble. The SMC setup allows for accurate sampling in spite of an approximate diffusion model and the finite stepsize used in the numerical solution of the stochastic process. Improved update strategies are also investigated. Results are presented for a $Z_2$ symmetric scalar field theory in 2 dimensions near its 2nd order phase transition.

\end{abstract}

\section{Introduction}\label{sec:introduction}

Investigating Quantum field theories (QFT), a very useful tool is
lattice discretisation.
The path integral is then approximated by finitely many integrals
over the variables
on the lattice. Analytical treatment is most likely
impossible, but Monte Carlo simulations 
provide a way to calculate the averages of observables
that we are interested in.

While Markov Chain Monte Carlo (MCMC) methods
are very successful, in some cases they are quite expensive.
For example, lattice QCD simulations are computationally demanding, as the cost of generating statistically independent gauge configurations are increasing rapidly as the continuum limit and physical quark masses are approached \cite{Schaefer:2010hu,Abbott:2022zsh}, but critical slowing down
is a problem generally, and it affects simulations of lattice theories
around a phase transition \cite{Wolff:1989wq,sokal1997monte}.
The rapid development of machine learning has prompted increasing interest
in using deep neural networks to reduce the computational cost of
lattice simulations.

Generative models in machine learning have become an active area of research across many fields of science \cite{deepgenerative}.
The aim is to sample a distribution
using the help of neural networks and machine learning. There
are many architectures such as energy-based models, variational autoencoders,
generative adversarial networks (GAN), autoregressive models, normalizing flows, etc. \cite{bond2021deep}, for QCD applications, see \cite{Cranmer:2023xbe}.
In field theories, GANs has been used to generate new samples and to reduce autocorrelation times \cite{Zhou:2018ill,Pawlowski:2018qxs,Wang:2020hji}.
Flow-based methods have seen more activity, with
attractive features such as the gauge equivariance and the cheap evaluation
of the Jacobian of the transformation \cite{Albergo:2019eim,Kanwar:2020xzo,Boyda:2020hsi,Albergo:2021vyo,Chen:2022ytr,Kanwar:2024ujc}.
Various modifications of the flow approach such as continuous and stochastic flow
have also been explored \cite{deHaan:2021erb,Caselle:2023mvh,Caselle:2022acb}.
However, it has also been established that the training costs increase rapidly
with the system size \cite{DelDebbio:2021qwf,Abbott:2022zsh,Komijani:2023fzy} and the method can suffer from
``mode collapse'', where
the model fails to explore separate peaks of the measure \cite{Nicoli:2023qsl,Kanaujia:2024zrq}.

  As a promising alternative, in this paper, diffusion models (DM) are used which have attracted
much interest over the past decade (For a general review, see \cite{yang2023diffusion}).
The application of diffusion models for lattice QFTs was first proposed in \cite{Wang:2023exq,Wang:2023sry}.
Its connection to path integrals was investigated in \cite{Hirono:2024zyg},
understanding the process in terms of cumulants was presented in \cite{Aarts:2024rsl}.
It has also been used in connection with the Complex Langevin equation \cite{Aarts:2025lpi} and for
U(1) gauge theory \cite{Zhu:2025pmw,Zhu:2024kiu}.
Recently, a self-learning variant has been investigated in \cite{Tomiya:2026tbc}, and it has been applied
to general gauge theories \cite{Vega:2025hgz,Aarts:2026zzr,Komijani:2026lan}.

In Section 2, a short introduction to diffusion models is given, while Section 3 introduces
the Sequential Monte Carlo (SMC) method. In Section 4, the new sampling method
using backward diffusion with ideas from SMC is introduced. In Section 5 the details of the setup of the
simulations and the numerical procedures are presented, while Section 6 contains the numerical results.
Section 7 offers conclusions.

\section{Diffusion models}

Diffusion models connect two ensembles, one of which is the target ensemble that we wish to sample, and one which is
tractable (e.g. a Gaussian ensemble). The two ensembles are morphed into each other using  diffusion processes.
In the forward process, configurations from the target ensemble are subjected to noise. For the sake of
concreteness, let's consider a scalar field theory of the fields $ \phi(x) $.
The forward process can then be described by a Langevin equation 
\bea
d \phi =  f(\phi,\tau)  d\tau + g(\tau) dw
\eea
which describes the evolution of the fields as the Langevin time $\tau$ increases, and $dw$ is the increment of a standard
Wiener process (or a vector of independent Wiener processes). The drift term, $ f(\phi,\tau) $ can be chosen arbitrarily, typically it is chosen to either vanish or to be linear in the fields.
The amplitude of the noise $g(\tau)$ can be  transformed into unity by redefinition of the Langevin time (also rescaling $f(\phi,\tau)$), it is kept
as a convenient tool to control numerical implementation. The forward process can be described by
$P(\phi,\tau)$,
which is the probability density of the fields dependent on the Langevin time $ 0 \le \tau \le \tau_{max}$.
Its time evolution is described by the Fokker-Planck equation
\bea \label{fokkerplanck}
    { \partial P( \phi,\tau) \over \partial \tau } = -{ \partial \over \partial \phi } ( f(\phi,\tau) P (\phi,\tau))
    + {g(t)^2 \over 2} \left( { \partial \over \partial \phi} \right)^2 P(\phi,\tau).
\eea
(Note that the more usual physics convention is to have $\sqrt{2} dw$ in the Langevin equation and consequently no
factor 1/2 in the last term of (\ref{fokkerplanck}).)
For vanishing drift, the variance of the fields
grows with time and eventually the initial variance (from the target ensemble) will be overwhelmed by the random noise coming from the diffusion process, and the distribution will be well approximated by Gaussians.

It turns out that the backward evolution
$ P_b(\phi,\tau')=P(\phi,\tau_{max}-\tau')$ is also a diffusion process, i.e. it is the solution of
a Fokker-Planck equation,
which corresponds to the backward Langevin equation \cite{anderson1982reverse} 
\bea  \label{backwardlangevin}
d \phi =  -f(\phi,\tau)  d\tau' +g(\tau)^2 \partial_\phi \ln P(\phi,\tau) d \tau' +  g(\tau) dw'.
\eea
Apart from the sign change of $f(\phi,\tau)$, the difference is the appearance of the new term $ g(\tau)^2 \partial_\phi \ln P(\phi,\tau) $ in the drift.
(Normally we parametrize the backward diffusion process also with $ \tau$, which then evolves from $ \tau_{max} $ to 0.)
Thus, to sample the target distribution, we
have to first sample the easy distribution, and let the configurations evolve
according to the backward diffusion process. At first sight this is not helpful as the drift term in the backward process requires the knowledge of $P(\phi,\tau)$, but the new drift term (aka. the 'score') can be fitted
(by a neural network)
if one has access to samples from the target distribution. This fit
is performed by following the forward diffusion process and minimizing a loss function \cite{vincent2011connection}. To generate samples
from the target distribution we thus sample the easy distribution and
solve the diffusion process (\ref{backwardlangevin})
with the replacement $ \partial_\phi \ln P(\phi,\tau) \rightarrow
s_\theta (\phi,\tau) $ where  $s_\theta (\phi,\tau)$
is called the score, and it is calculated by a neural network,
$\theta$ representing the parameters of the network.

\section{Sequential Monte Carlo and diffusion processes}

In Sequential Monte Carlo (SMC) \cite{doucet2000sequential,del2006sequential,andrieu2010particle}, one seeks to obtain averages using a number
of probability measures $ \pi_t(\phi) $ with $t=0 \dots N_\tau$, where $\phi$ is an element of some manifold $V$ (one can think of e.g. $ V= \mathbb{R}^n $ or $ G^n$ with some Lie-Group $G$). Typically the initial distribution
 $\pi_0(\phi)$ is easy to sample, for example it is a
Gaussian distribution or it can be dealt with efficiently using Markov-chain Monte Carlo
methods. One desires to sample the target distribution $ \pi_N(\phi) $.

The average of an observable $ F(\phi)$ is defined by 
\bea
 \langle F \rangle_t = { 1 \over Z_ t}  \int_V d\phi \pi_t(\phi) F(\phi) 
 \eea
 with $ Z_t = \int_V \pi_t(\phi) d\phi $, and $ \int_V d\phi $ denotes an integral over the whole manifold $V$.

Averages are estimated using a cloud of ``particles'' $ \phi_i$ with $ 1 \le i \le M $ with weights $w_i$.
Initializing $\phi_i$ with random samples distributed as $ \pi_0(\phi) $, and setting
$w_i = 1/M $, we can easily see that 
\bea \label{particleavr}
 \langle F \rangle_0 = { \sum_{i=1}^M F(\phi_i) w_i \over \sum_{i=1}^M w_i }
\eea
indeed gives the average as defined above for the initial distribution $ \pi_0$.
To estimate averages for $ \pi_t(\phi) $ with $t>1$,
we introduce a new notation $ \phi_i =\phi_i^{(0)} $, $ w_i = w_i^{(0)} $, 
and we update $ \phi_i^{(t)} \rightarrow \phi_i^{(t+1)} $ as well as 
$ w_i^{(t)} \rightarrow w_i^{(t+1)}$ such that
\bea \label{particleavrt}
 \langle F \rangle_t = { \sum_{i=1}^M F(\phi_i^{(t)}) w_i^{(t)} \over \sum_{i=1}^M w_i^{(t)} }
\eea
remains correct for $ \langle F \rangle_t$.
We have several update strategies, the simplest is:
\bea
  \phi_i^{(t+1)} = \phi_i^{(t)} , \quad w_i^{(t+1)} = w_i^{(t)} { \pi_{t+1}\left(\phi_i^{(t+1)}\right) \over \pi_t\left(\phi_i^{(t)}\right) }   
\eea
which keeps the particles fixed and only updates the weights according to the change of the measure. Again it's easy to prove that this update strategy ensures that (\ref{particleavrt}) is correct, however, this strategy can still be impractical as the magnitude of the weights can vary differently, and if the average is dominated by a few $i$  with the largest $ w_i^{(t)} $ values, then the estimate for the averages will have huge variance. To monitor this behavior, we introduce the observable called effective sample size (ESS)
\bea
\textrm{ESS}= {   \left(\sum_{i=1}^M w_i ^{(t)} \right)^2        \over \sum_{i=1}^M (w_i^{(t)}) ^2  }.   
\eea
If all weights are equal then ESS=$M $, whereas if one $ w_i $ is vastly larger
than the others then  ESS $\approx 1 $.
Using this observable the usefulness of an update strategy can be judged by the decrease of the ESS. Below some resampling
strategies are mentioned to deal with the ESS decreasing too much.

A second general update strategy is to use random updates given by the
probability density $ K_t(\phi^{(t)},\phi^{(t+1)}) $. The update is given by
\bea
  \phi_i^{(t+1)} = \psi_i, \quad w_i^{(t+1)} = w_i^{(t)}  { \pi_{t+1}\left(\psi_i \right) \over \pi_t\left(\phi_i^{(t)}\right) }   { L_t(\psi_i,\phi_i^{(t)}) \over K_t(\phi_i^{(t)},\psi_i)   }
\eea
where $\psi_i$ is a random sample which, for a given $ \phi_i^{(t)}$, is
chosen according to the probability density $ K_t(\phi_i^{(t)},\psi) $ (called also as the {\it kernel} of the update).
$ L_t(\psi,\phi) $ is a backward kernel, describing a random process which proceeds backward, giving $\phi$ for a given $\psi$. Its kernel, $L_t(\psi,\phi)$, can in principle be chosen arbitrarily, as long as 
\bea
 \int_V d\phi L_t(\psi,\phi) =1
\eea
is satisfied. Irrespective of the actual $L_t$, the averages (\ref{particleavrt}) remain correct.
The optimal $L_t(\psi,\phi)$ to be used in practice is the one that minimizes the variance of weights, or equivalently, maximizes ESS. 

If the ESS decreases too much during the process (in practice it is less than e.g. $0.5 M$), we
can regenerate the particle cloud by {\it resampling}.
We select $M$ random integers $ 1\le i' \le M $ according to the current
probabilities $ w_i $ (with replacement), and use $ \phi_{i'} $ as the new particles.
Finally, we update all weights to $1/M$. This construction preserves the averages, and maximizes the $ESS$ observable, however, it introduces correlations in the dataset as one particle can be selected multiple times if its weight is large. To get rid of the correlations, one might use MCMC updates
using the current $ \pi_t(\phi)$ measure before continuing the update
(also called as {\it rejuvenation} of the sample).
See e.g. \cite{del2006sequential} for advanced resampling strategies (reducing the variance of the estimators of observables)
such as ``systematic resampling''.

 Note that, using (\ref{particleavrt}), since the weights are now also random variables, there is
an $O(1/M)$ bias, as we are calculating $\langle A/B \rangle $ instead
of $ \langle A \rangle / \langle B \rangle  $, where $A$ and $B$ are
the numerator and denominator on the right hand side of (\ref{particleavrt}).
(This remains true if we are using resampling, see in \cite{moral2004feynman}.)
This bias is,
however, overwhelmed by the statistical errors of size $ O(1/\sqrt{M})$, when we
estimate $\langle F \rangle$  with (\ref{particleavrt}).

Finally, let's discuss SMC as it relates to a diffusion process, given by the
time dependent density $ P(\phi,\tau) $, corresponding to the Langevin process
\bea \label{smc-langevineq}
 d\phi = a(\phi,\tau) d \tau + g(\tau) d w. 
\eea
A discretised diffusion process gives us the distributions $ \pi_t(\phi) = P(\phi,\tau_{max} t/N_\tau ) $, so
we can understand a diffusion process in terms of SMC.

For the update of weights, we need to calculate the forward kernel and
we have to choose a backward kernel.
The forward kernel $K(\phi,\psi)$ is calculated from the update of the
discretised version of the Langevin equation (\ref{smc-langevineq}),
see details below.
For the backward kernel $L(\psi,\phi)$, the natural choice here is the diffusion update corresponding
to the backward diffusion process of the same $P(\phi,\tau)$ distribution,
as discussed in the previous section.
In this case the update will be given by (dropping the particle index  $i$ and time index $t$)
\bea
\phi \rightarrow \psi , \quad     w \rightarrow w
{ P(\psi,\tau+\Delta \tau) \over P(\phi,\tau) } { L(\psi,\phi) \over K(\phi,\psi) }
\eea
The backward diffusion kernel is given by
\bea
 L(\psi,\phi)  = P( \phi_\tau=\phi | \phi_{\tau+\Delta\tau}=\psi ) = { P(\phi,\tau) K (\phi, \psi ) \over P( \psi,\tau+\Delta\tau) }
\eea
and the weight update becomes $  w \rightarrow w $, which is obviously
the optimal choice, maximizing ESS.

To calculate the kernels, we use the first-order discretised update
of the Langevin equation (using $\Delta \tau= \tau_{max}/ N_t $),
 which is the Euler-Maruyama update
\bea
 \phi^{(t+1)} = \phi^{(t)} + a(\phi,t \tau_{max}/ N_t )  \Delta \tau + g(\tau) \eta \sqrt{ \Delta \tau}.
\eea
This gives 
\bea
K_t(\phi,\psi) =\exp \left( - { (\psi-\phi -a(\phi,\tau_{max} t /N_t) \Delta \tau )^2 \over   2 g(\tau)^2 \Delta \tau } \right),
\eea
where we have dropped irrelevant constants not depending on $\phi$ or $\psi$.
Below, we will use SMC for calculating averages during the backward
diffusion process, such that the drift term is given by
$ a(\phi,\tau) = g^2(\tau) \partial_\phi \ln P (\phi,\tau) - f(\phi,\tau) $, where $f(\phi,\tau) $ is the drift term that was used during the noising process.
Now $L(\psi,\phi)$ describes the backward of the backward process, that is the
forward, noising process. $L(\psi,\phi)$ is then easily calculated as
\bea
L_t(\psi,\phi) =\exp \left( - { (\phi-\psi - f(\phi,t \tau_{max}/N_t) \Delta \tau )^2 \over   2 g(\tau)^2 \Delta \tau } \right),
\eea
where we have assumed a linear (or zero) $ f(\phi,\tau) $.
For the discretised system, the weight update will be an exact identity
only in the limit $ \Delta \tau \rightarrow 0 $. For a finite but low stepsize
the weights will drift away from their initial value and the ESS will slowly decrease.
These weights, however, keep the averages accurate for any finite stepsize.

The forward and backward kernels are particularly easy to calculate above for the
Euler-Maruyama update, for improved updates the expression for $K(\phi,\psi)$ can be more complicated,
as it typically involves a determinant of some Jacobian, which is numerically expensive, see Section 5 for details.

\section{SMC aided Diffusion models}
\label{setsec}

We seek to investigate theories given by some action $S(x)$ with variables $x$ in some manifold $V$.
The aim of this paper is to provide an algorithm for Monte Carlo estimation of
 exact averages for expectation values 
\bea
 \langle F \rangle = { \int_V d\phi F(\phi) e^{-S(\phi)} \over \int_V d\phi e^{-S(\phi)} }
\eea
using diffusion models.

Using samples from the theory distributed according to the measure $ e^{-S(\phi)}$ (which we get from some potentially
expensive source),
we train a neural network to approximate the drift terms (i.e. the score) in the backward diffusion process.
We can thus create new configurations of the theory using the backward diffusion process \cite{Wang:2023exq,Zhu:2025pmw}.
However, since the neural network only approximately reproduces the score, the distribution
of the configurations will only be approximately equal to the desired measure $ e^{-S(\phi)} $.
One therefore has to amend this setup with some kind of extension to make sure the distributions are exact.

Here, a new approach is proposed: we use Sequential Monte Carlo to make sure that we are sampling from a sequence of known distributions such that the last one is given by the measure $ e^{-S(\phi)} $.
This sequence is naturally built in to the backward diffusion process, giving
the distributions $ P(\phi,\tau) $ for $ 0 \le \tau \le \tau_{max} $.
(Using the convention
that $ 0 \rightarrow \tau_{max} $ gives the forward process and $ \tau_{max}  \rightarrow 0$ is the backward process, and $\tau_{max}=1$ is set in the numerical tests.)
First of all,
we use an Energy-based representation of the process, such that the neural network learns
$ R(\phi,\tau) $ which is, up to a constant shift, the negative logarithm of the distribution of the fields during the
process as $\tau$ goes from $\tau_{max}$ to 0.
For the SMC setup we take (note that we don't need to normalize the distributions for SMC)
\bea
\pi_0(\phi) = G(\phi), \quad 
\pi_t(\phi) = e^{-R(\phi,(1- t /N_\tau) \tau_{max})} \textrm{ for }  \ 1 \le t < N_\tau, \quad 
\pi_{N_\tau}(\phi) = e^{-S(\phi) }
\eea
where $G(\phi)$ is the idealized initial distribution for the backward diffusion process (Gaussian for non-compact manifolds, uniform distribution for compact ones).
If the neural network would learn the distributions perfectly, in the limit
of infinite $\tau_{max}$, the backward diffusion naturally follows
the $ \pi_t$ distributions, so all weights in the sequential
Monte Carlo method should remain constant. In practical use however, the variance of the weights
is increasing (and ESS is decreasing) due to the following effects:
first, for finite $\tau_{max}$ the initial distribution is typically well approximated by a Gaussian/uniform distribution, but it's not exactly that.
Second, when numerically following the backward diffusion process, we inevitably have some finite-stepsize effects.
Third, the distribution learned by the neural network is only an approximation to the backward diffusion process.
Following the evolution of the weights in the SMC process eliminates all three sources of errors, but the effects contribute to a decreasing ESS.
All effects can be decreased with some effort: first,  one needs to make sure that the noised
configurations are close to Gaussian (uniformly) distributed, which is easily
arranged by choosing a $ g(\tau)$ function which increases the total noise
amplitude in the forward process.
Second, a sufficiently
small Langevin stepsize and/or improved updates are to be used in the denoising process, and
third, a better neural network architecture and more learning should improve
the approximation of the distributions.
In practice, if these requirements are not satisfactorily met,
then the ESS decreases quickly, and the variance of the results will be larger.


We will refer to this setup as DM-SMC (Diffusion Model - Sequential Monte Carlo).
The costs of following the evolution of the weights
are typically negligible in comparison with the update for the fields,
but they let us cancel the finite stepsize effects,
as well as keep the target distribution exact, in
spite of the diffusion model, as being fitted by a neural network, being an inexact
approximation of the diffusion process between the two ensembles.

Existing setups for generating samples using diffusion models are easily upgraded to DM-SMC, provided they use the energy-based representation of the score:
the generation of $M$ configurations has to progress in parallel, the weights
must be followed, the ESS must be monitored and a resampling has to take place
if the ESS decreases too much. Apart from calculating the ESS and the
resampling there
is no communication between
the parallel threads of the generation.


In DM-SMC, the rejuvenation updates can be conveniently used from the Langevin equation
\bea 
d \phi = {1\over 2}  g(\tau)^2 \partial_\phi P(\phi,\tau) d \tau +  g(\tau) dw,
\eea
which is just the backward diffusion process with the drift term halved, and
the $ \tau$ argument of the drift and noise terms kept fixed during the process (and dropping
$f(\phi,\tau)$ in case it is nonzero).
This process has the stationary distribution $ \pi_t(\phi) = P(\phi,\tau) $, as can be easily seen.
(This is the usual recipe to set up a Langevin equation for the measure $ \exp ( -S(\phi)) $:
 $ d\phi = - S'(\phi) d\tau + \sqrt{2} d w $, used here with a rescaled Langevin time).
To make sure that the distributions remain exact (in spite of the finite Langevin step),
an accept-reject step is performed after each Langevin update
(which makes the process a Metropolis-adjusted Langevin Algorithm (MALA)
\cite{roberts1998optimal}).

\section{Simulation setup}

To solve the stochastic differential equation 
\bea \label{generalSDE}
 d \phi = a( \phi,\tau) d \tau + g(\tau) dw
\eea
numerically, one can employ the Euler-Maruyama update
\bea
 \phi(\tau+\Delta \tau) = \phi(\tau) + a(\phi,\tau) \Delta \tau + \eta_\tau g(\tau) \sqrt{\Delta \tau},
 \eea
 where $ \eta_\tau $ is a Gaussian random variable with zero mean and
 unit variance (or a vector of such independent random numbers). This update is correct to first order in $ \Delta \tau$,
 and has the kernel density
 \bea
 \ln K(\phi,\psi) = - {\eta^2\over 2} +  \textrm{const.}
 \ \ \textrm{ with } \ \ \eta = {\psi-\phi-a(\phi,\tau) \Delta \tau \over g(\tau) \sqrt{\Delta \tau} }
 \eea
 A (weak) order 2 update to solve (\ref{generalSDE}),
 using $ \phi'=\phi(\tau+\Delta \tau) $
 is given by
 \bea \label{improvedupdate}
  \phi'  = \phi(\tau) + { a(\phi,\tau)+ a(\phi',\tau+\Delta \tau) \over 2}    \Delta \tau + \eta_\tau
 { g(\tau) + g(\tau+\Delta \tau) \over 2} \sqrt{\Delta \tau},
 \eea
 This equation is implicit, to solve it one can e.g. calculate a first
 approximation to $ \phi'$ using the Euler-Maruyama update, then
 iterate (\ref{improvedupdate}) until convergence (typically 2-4
 iterations are enough to reach machine precision).
 The update kernel is now more complicated as it also involves
 the Jacobian of the drift term, $Da(\phi,\tau)$
 \bea
 \ln K(\phi,\psi) = - {\eta^2\over 2} +
 \ln \left| \det \left( 1 - {\Delta \tau \over 2 } Da(\phi,\tau +\Delta \tau ) \right)  \right|+
 \textrm{const.}
 \ \ \textrm{ with } \ \ \eta = {\psi-\phi-(a(\phi,\tau)+a(\psi,\tau+\Delta \tau))) \Delta \tau/2 \over
   \sqrt{\Delta \tau} (g(\tau)+g(\tau+\Delta \tau))/2 }
 \eea
 Calculating the determinant is prohibitively expensive for lattice models
 (except for small volumes). For small enough $\Delta \tau $ one
 can approximate the logarithm of the determinant (which is positive in this case) with
 \bea \label{detexpand}
 \ln \left| \det \left( 1 - {\Delta \tau\over 2 } Da(\phi,\tau + \Delta \tau) \right)  \right|
 =  - { \Delta \tau \over 2 } \textrm{Tr} D a(\phi,\tau+\Delta \tau) + O(\Delta \tau ^2)
 \eea
 The trace can be efficiently approximated using $ N_v $ noise vectors $ v_k,
 \ \ k = 1 \dots N_v $,
 which satisfy $\langle v_k \rangle = 0 $ and $ \langle v_k v_j^\dagger \rangle = I \delta_{jk}$
 with $I$ the identity matrix:
 \bea \label{traceapprox}
 \textrm{Tr} D a(\phi,\tau) \approx {1\over N_v } \sum_k v_k^\dagger D a(\phi,\tau) v_k.
 \eea
 This approximation however reintroduces $\Delta t$ dependence and an
 extrapolation $ \Delta t \rightarrow 0 $ is needed.
 Whether the first order update which gives exact weights and no extrapolations are needed or the second order update
 giving approximate weights and the extrapolation afterwards is cheaper needs to be investigated
 for the particular application one has.

 To investigate how this proposal performs in practice,
 following \cite{Wang:2023exq}, a real scalar field in Euclidean space-time of $d$ dimensions is considered with the action
\bea
S = \int d^d x \left( (\partial_\mu \phi_0)^2 + {1\over 2} m_0^2 \phi_0^2 + {\lambda_0\over 24} \phi_0^2 \right).
\eea
Simulations of the scalar theory, discretised on a cubic lattice with lattice action
\bea
S = \sum_{x} \left[ -2 \kappa \sum_{\mu=1}^d \phi(x) \phi(x+\hat \mu) + (1 -2 \lambda ) \phi(x)^2 + \lambda \phi(x)^4 \right], 
\eea
are carried out in $d=2$ dimensions, where the scalar field has been rescaled using $ \phi = a^{d/2-1} (2 \kappa) ^{-1/2} \phi_0 $ with the lattice spacing $a$,  $\phi(x+\hat\mu)$ is the neighbor field in positive $\mu$ direction,
$ \kappa $ is the hopping parameter which is calculated from the bare mass parameter using the formula
\bea
 (a m_0 )^2 = { 1 - 2 \lambda \over \kappa} -2d 
\eea
and $ \lambda$ is calculated as $ \lambda  = 2 \lambda_0 a^{ 4-d} \kappa /6 $.
 The main observables are the  action density  $ s = S/(N_x N_y) $ with $N_x$ and $N_y$,
 the size of the periodic lattice and 
 the average field (aka. ``magnetization'')
 \bea
 \Phi = {1\over N_x N_y } \sum_{x} \phi(x).
 \eea
This scalar theory has a second order phase transition
for each $\lambda$ value as $\kappa$ is increased, corresponding
to the spontaneously broken $Z_2$ symmetry of the theory. 
For simulation the parameters are chosen as $\kappa=0.27$ and $ \lambda=0.022$,
which are close to the critical values, such that the histogram of the magnetization at this point shows a double peak structure. The reason for
investigating this point is that a phase
transition region is typically the hardest to deal with for MCMC methods
as well as for generative methods. 

In this study, for noising and denoising the variance expanding diffusion processes are used, which means $ f=0$ and $ g(\tau)= \sigma^{\tau} $, 
giving
\bea
 \Sigma^2(\tau) = \int_0^\tau g^2(t) dt = { \sigma ^{2 \tau} -1 \over 2 \ln \sigma },
\eea
and thus the total variance that is added to the fields by the diffusion process is $ \Sigma^2(1) = ( \sigma^2 -1) / (2 \ln \sigma )  $. Unless otherwise specified, $\sigma=25$ is used below.

 Finally, some details about the neural networks used in this study:
 as written above we use the ``energy-based'' setup (instead of the ``score-based'') as
 for the SMC method a probability measure is needed for $ 0 \le \tau \le 1 $ during
 the denoising process. This is conveniently supplied by the ``energy''.
 We thus need a neural network that has as inputs the field configuration $ \phi(x) $
 as well as $ \tau$ and it outputs one scalar $ R(\phi,\tau) $ such that
 the probability measure of the fields
 at time $\tau$ is $ \exp(-R(\phi,\tau))$. The result of the network
 should be invariant to translations of the
 lattice (periodic boundary conditions
 are used), as well as to rotations/mirrorings that leave the lattice
 invariant. A natural choice that satisfies these
 requirements is a convolutional neural network. The network is built out
 of several (2-4)
 convolutional layers, each of which is built of a convolutional
 kernel of 3x3 size,
 and local mixing layer afterwards. (Each linear layer is followed by an activation layer).
 After the last convolutional layer a final linear layer calculates a linear combination
 of all the channels in the last convolution layer and finally a sum over the lattice is performed
 to calculate the final result.
 To allow the network to fit the dependence on $\tau$, before the
 first layer the configuration is extended to a second channel, which is a constant
 value of the $ \Sigma(\tau)  = \sqrt{\int_0^\tau g^2(t) dt} $, which is the
 square root of the variance of the noise that is added to the fields
 until time $\tau$  during the noising process. The inner convolution layers work with
 up to 4-16 channels.
  The invariance of the convolution kernels under lattice rotations/mirroring
 is not enforced in this study. The swish activation function is used.
 
 To train the network the loss function is built out of two terms $ L = \alpha L_1 + (1-\alpha) L_2  $, 
 where the first term is \cite{vincent2011connection,Wang:2023exq,Zhu:2025pmw} 
 \bea 
 L_1 =  \sum_{(\phi_0,\tau,\eta)} \left| \Sigma(\tau) s_\theta(\phi_\tau,\tau) + \eta \right| ^2
 \eea
 Where $s_\theta(\phi_\tau,\tau)= - \partial_\phi R(\phi,\tau) $ is the score that the network gives for the configuration $\phi_\tau$ at time $\tau$ and $\theta$ represents the current parameters of the network.
 The sum is over the triples $ (\phi_0 , \tau , \eta )$ with a configuration $\phi_0$ that is sampled from
 the target ensemble, $ \tau $ a random time $ 0 \le \tau \le \tau_{max} $, and
 a Gaussian random vector $\eta$ with zero mean and unit variance which has the
 same dimension as $\phi$.
 $ \phi_\tau$ is created by using $\eta$ to noise the configuration $\phi_0$ (for time $\tau$):
  $ \phi_\tau = \phi_0 + \eta \Sigma(\tau) $.
 This loss function is in principle is enough to let the network fit the noising/denoising process. In
 addition,  to have more control over the learning process, the second term
 \bea
 L_2 = (S(\phi) - R(\phi,\tau=0))^2
 \eea
 is also used, which makes sure that at $ \tau=0$ the ``energy'' of the denoising process
 coincides with the action we wish to sample. In the training process typically an $\alpha$ value
 of 0.5-1.0 is used.

\section{Results}
\label{ressec}

To provide benchmark values, estimation of the averages using Hybrid Monte Carlo simulations \cite{Duane:1987de} are also performed.

In some cases, resamplings (using systematic resampling) of the particle cloud are used when the ESS decreases to 0.5 times its initial value.
After resampling, $ N_\textrm{rejuv} =5 $ Langevin steps are used with drift terms calculated from the
current measure $ R(x,\tau) $ to decrease correlations (see in section 4). We use an accept-reject step
to ensure exact distributions, (MALA algorithm, see \cite{roberts1998optimal}),
and the stepsize of this update is set such that
the acceptance rate is around $ \approx 0.6 $, which in practice means $\Delta t_\textrm{rejuv} > 0.025$.
The size of the particle cloud is chosen to be in the range $ M = (1-16) \cdot 10^3 $ in the simulations below.

\begin{figure}
\begin{center}
  \epsfig{file=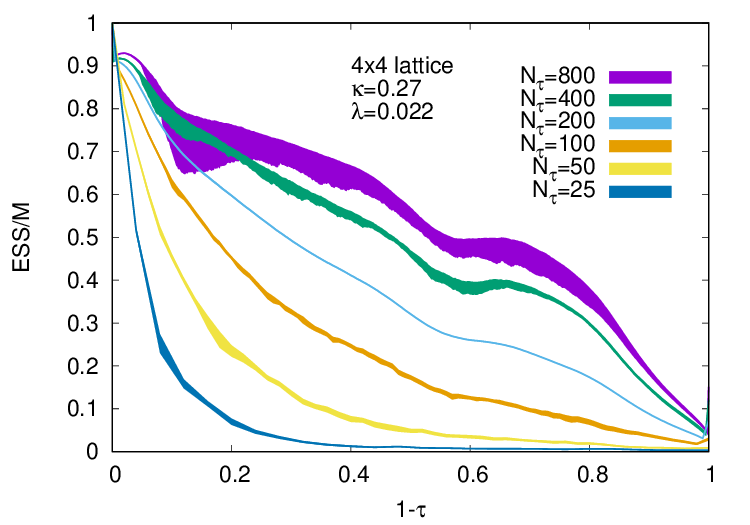, width=8.5cm}
  \epsfig{file=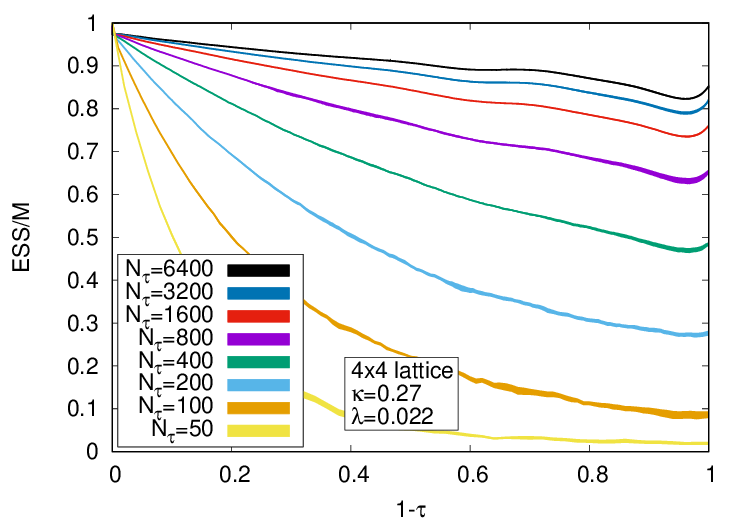, width=8.5cm}
  \caption{ \label{ess4x4}
    The effective sample size over the number of total samples on a 4x4 lattice for  
    $ \kappa =0.27 $ and $ \lambda=0.022 $ and various Langevin stepsizes, as indicated by the total number of steps.
    The first order Euler-Maruyama update is used.
    On the left, the learning of the model was stopped 200 learning steps. On the right, the learning was allowed to converge using $\approx 1000$ learning steps, and the parameters of the model were averaged for a further 1200 steps.
    No resampling was used.
}
\end{center}
\end{figure}
To investigate the behavior of the effective sample size (ESS), simulations
on $4^2$ lattices without resampling are performed.
In Fig.~\ref{ess4x4} the ESS is shown for various stepsizes in the backward process. Decreasing the Langevin stepsize increases ESS, however, one notices that
decreasing the stepsize can only increase ESS up to a point, which is determined by the quality of the
fit of the distributions by the neural net. Note that the very first point of the curve is at 1.0,
after which the curve jumps to a smaller value. This is a consequence of setting $ \pi_0 $ to the initial
Gaussian and not to $ R(\phi,1) $. (Also, we have an imperfect fit of the diffusion process by the neural network). The size of the jump indicates that to describe the distribution
of the noised configurations, we still need some corrections to the idealized Gaussian distribution.
This jump can be improved upon using the $ \sigma$ parameter of the diffusion process, such
that a larger $ \sigma$ value means that the total amplitude of the noise added to the
field is larger, and thus the Gaussian will be a better approximation. A similar jump is to be observed
at the last point of the curve, corresponding to $\tau=0$, as $ \pi_{N_\tau} = e^{-S} $ is set by hand.
The jump in this case can also be upwards, this is dependent on the particular fit
the neural network gives.
In the right panel of Fig.~\ref{ess4x4}, we show again the ESS for a model which was fitted much more carefully:
the learning process was longer and the parameters of the model are averaged after learning
has converged and 
the model is fluctuating around a minimum of the loss function. One observes that for equal stepsize,
the ESS for the better model decreases much more slowly. To obtain a nearly flat ESS curve, however,
the stepsize has to be quite small.

\begin{figure}
\begin{center}
  \epsfig{file=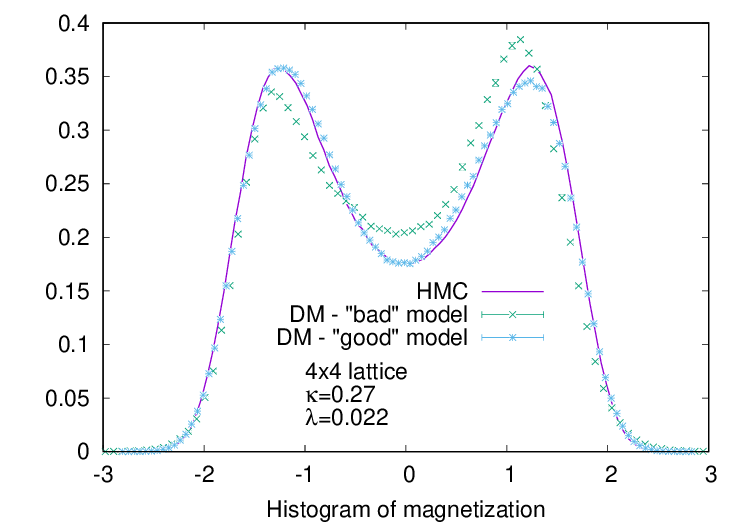, width=11.5cm}
  \caption{ \label{magnethist-nosmc}
    Histogram of the field average $\Phi$
    for a 4x4 lattice for 2d scalar field theory using
    $ \kappa =0.27 $ and $ \lambda=0.022 $.
    Configurations are sampled using denoising (without SMC setup) with the
    two models in Fig.~\ref{ess4x4} and $ N_\tau =200 $ steps for the ``bad'' model
    and $ N_\tau=200 $ steps with improved update steps for the
    ``good'' model.
    Results from an HMC simulation are shown as well.
}
\end{center}
\end{figure}
In Fig.~\ref{magnethist-nosmc} the histogram of the magnetization is shown using the two models
in Fig.~\ref{ess4x4} and denoising without the SMC setup. We see that both models
give histograms which are not exact, where the ``good'' model and an improved
update step leads to a better approximation, as expected.

\begin{figure}
\begin{center}
  \epsfig{file=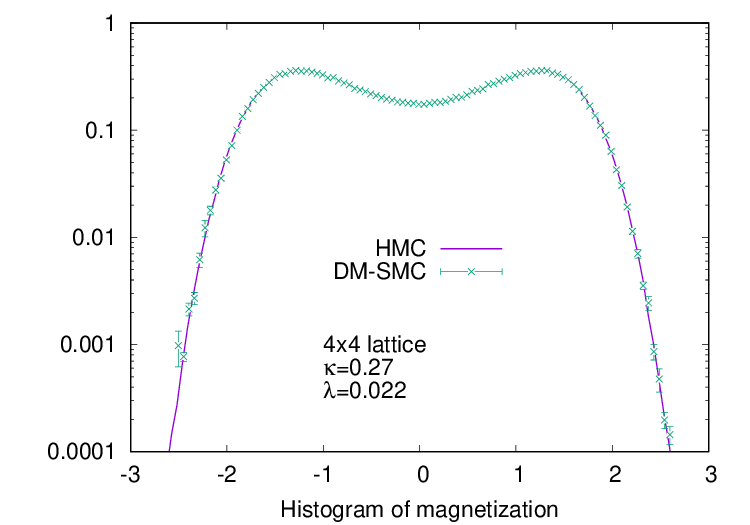, width=8.5cm}
   \epsfig{file=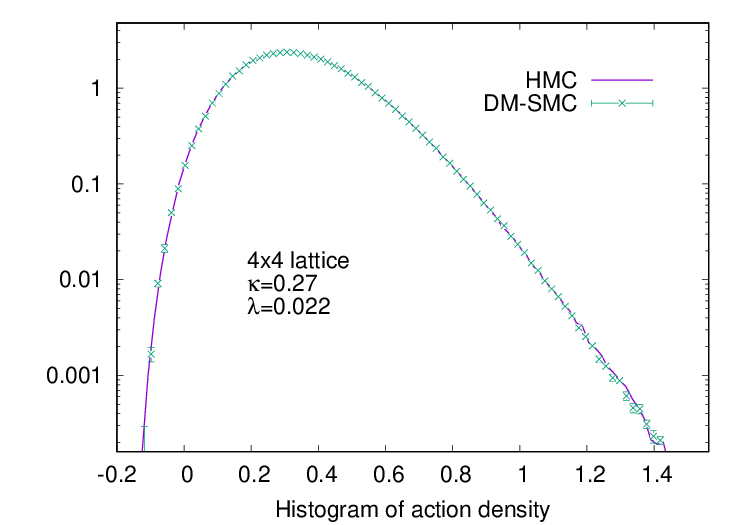, width=8.5cm}
  \caption{ \label{magnethist}
    Histogram of the field average $\Phi$ (left panel) and histogram for the action density $s$ (right panel) for a 4x4 lattice for 2d scalar field theory using
    $ \kappa =0.27 $ and $ \lambda=0.022 $. Measurements using the DM-SMC setup
    use the model corresponding to the left panel of Fig.~\ref{ess4x4}, $ N_\tau=200$, and one
    resampling at the end of the denoising process.
    Results from an HMC simulation are shown as well.
}
\end{center}
\end{figure}
In Fig.~\ref{magnethist}, the histogram of the magnetization as well as the histogram
of the action average are shown, using DM-SMC, the SMC aided denoising
setup. We see perfect agreement (within statistical errors) with the benchmark values
from an HMC simulation, even though the not accurately fitted model is used
from the left panel of Fig.~\ref{ess4x4}, with 200 first-order Langevin steps in the
denoising process. (Fig.~\ref{magnethist-nosmc} shows how the histogram would look without the SMC setup).
Only one final resampling is used at $ \tau=0$, the end of the
denoising process, where the ESS/$M$ average for these parameters is $ .0716 \pm .0006$ (Note that there is small jump
of the ESS upwards in the last update step which is hardly visible in Fig.~\ref{ess4x4}). $ O(10^3) $ independent runs, each of which uses $M=8192$ are
averaged to
decrease statistical errors.
If we use a better fitted model and/or smaller Langevin stepsize, the
results remain consistent with the exact results, and statistical errors decrease as the ESS is higher.

\begin{figure}
\begin{center}
  \epsfig{file=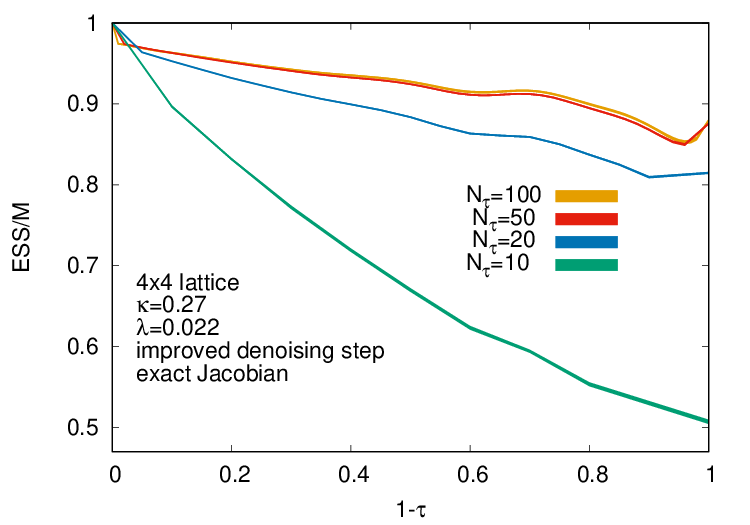, width=8.5cm}
  \epsfig{file=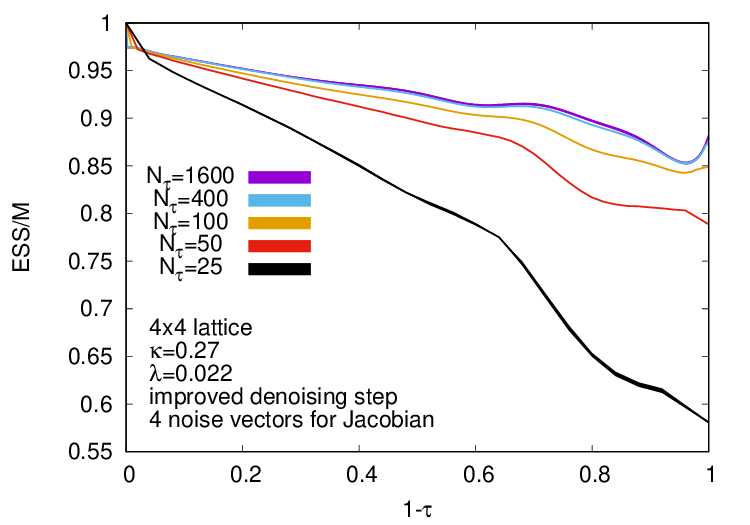, width=8.5cm}
  \caption{ \label{improvedESS}
    The ESS is shown for an improved denoising update for various
    Langevin stepsizes, as indicated. In the left panel,
    the determinant of the Jacobian is calculated exactky. In the right panel,
    the denterminant of the Jacobian is approximated using 4 noise vectors with
    the formulas in eqs. (\ref{detexpand}) and (\ref{traceapprox}).
}
\end{center}
\end{figure}
\begin{figure}
\begin{center}
  \epsfig{file=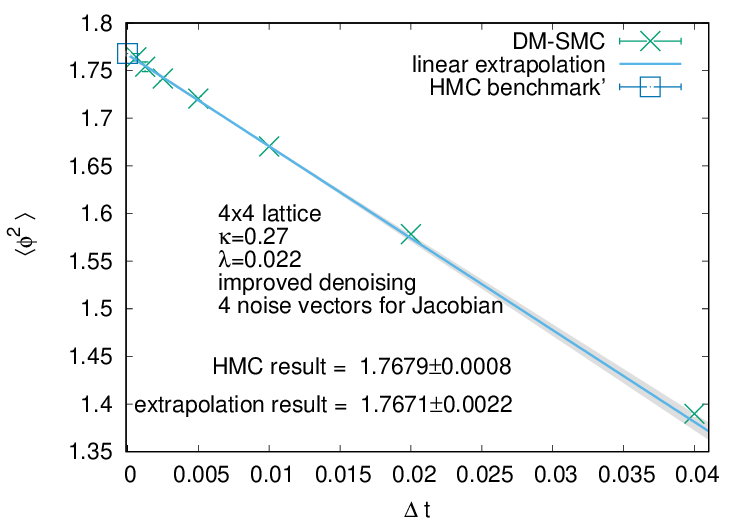, width=8.5cm}
  \caption{ \label{improved-extrap}
    The average of $ \phi^2 $ is shown for various stepsizes,
    as calculated from DM-SMC using the 2nd order update and the
    approximation of the
    Jacobian with 4 noise vectors.
    Extrapolation to zero stepsize, using stepsizes $\le 0.01$, and an HMC benchmark value is also shown.
}
\end{center}
\end{figure}

Next, the improved update is investigated. In the left panel of Fig.~\ref{improvedESS} we show the ESS
for the ``good'' model and the improved update (\ref{improvedupdate}), where the Jacobian
of the update is calculated exactly for updating the weights.
One observes a much slower decay of the ESS as compared to the first order update, such that
already $ \Delta \tau=0.1$ presents an ESS/$M$ which is still above 0.5 at the end of
the denoising process, therefore only one resampling at the end is necessary.
 $ \Delta \tau=0.02 $ already saturates the
improvement of the 
ESS decay (which is then determined by the inaccuracy of the diffusion model).
For such a small lattice the Jacobian 
calculation is still feasible. The cost of calculating the determinant
grows quickly with the lattice size, and this approach quickly becomes infeasible.
Therefore we test an approach which remains usable for larger lattices: 
in the right panel of Fig.~\ref{improvedESS}, 
the Jacobian in the weight update is approximated using the 
expansion of the determinant (\ref{detexpand}) and 4 noise vectors to estimate the
trace in (\ref{traceapprox}).
Also for the approximated update the decay remains slow, and here $ \Delta \tau=0.0025=1/400$, is enough to
reach the slowest ESS decay.
However, since the weight update is approximated, results have a bias for finite Langevin stepsizes, and the extrapolation to zero stepsize must be carried out, as
illustrated in Fig.~\ref{improved-extrap}.

\begin{figure}
\begin{center}
  \epsfig{file=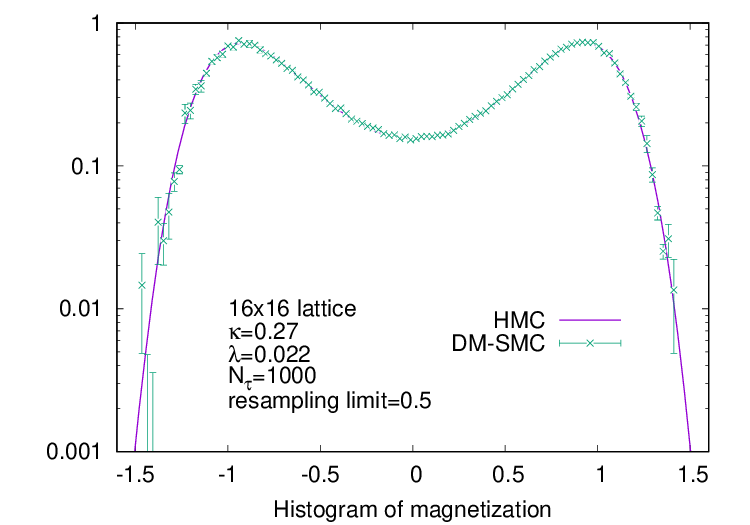, width=8.5cm}
   \epsfig{file=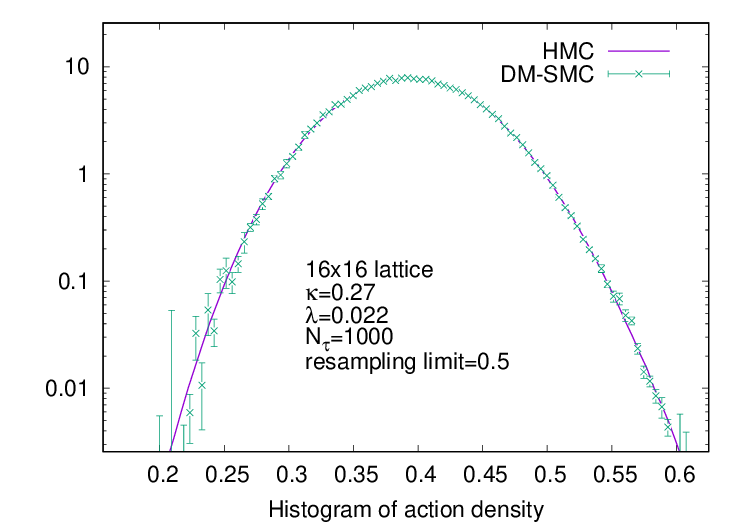, width=8.5cm}
  \caption{ \label{magnethist-16}
    Histogram of the field average $\Phi$ (left panel) and histogram for the action density $s$ (right panel) for a $16^2$ lattice for 2d scalar field
    theory using
    $ \kappa =0.27 $ and $ \lambda=0.022 $.
    The diffusion process uses $\sigma=50$. Denoising is done
    using $ N_\tau=1000$, and systematic resampling when the ESS/$M$
    decreases below 0.5.
    Results from an HMC simulation are shown as well. 
}
\end{center}
\end{figure}
Finally, the dependence on the system size is considered. The procedure was repeated on a $16^2$ lattice. 
As the system size increases, the decay of the ESS gets faster, therefore we use resamplings during the denoising process, whenever
the ESS/$M$ (with $M$ the number of samples in the particle cloud) decreases below 0.5. In practice, for the model
and parameters we use, this corresponds to 10-11 resamplings during the denoising process, depending on the random seed used.
In Fig.~\ref{magnethist-16} the histogram of the magnetization as well as the histogram of the action density are shown.
140 independent runs are averaged, each of which uses $M=8192$.
We see agreement within statistical
errors with the benchmark values calculated with the HMC algorithm.

\section{Conclusions}
\label{concsec}

 In this paper, a new sampling method called DM-SMC (Diffusion Model - Sequential Monte Carlo)
 is proposed which samples ensembles defined in terms of an action.
 We first train a diffusion model using samples from the ensemble (which we have to
 create using a potentially costly method). The diffusion model can be
 a good approximation to the exact diffusion process, but it will remain inexact.
 The SMC setup allows for accurate sampling in spite of an approximate diffusion model and finite
 stepsizes in the diffusion process.
 The method is based on following the ensemble using a cloud of ``particles'', i.e.
 configurations of the ensemble, such that their weight is also kept track of,
 and can change during the denoising process, in the spirit of the Sequential Monte Carlo method.
 The results are exact up to $O(1/M)$ corrections with the number of particles $M$,
 but this bias is overwhelmed by statistical errors of $ O(1/\sqrt{M})$. 
 
 This proposal is tested on a scalar field theory which has a 2nd order phase transition, and
 we tune to a point in the parameter space which is close to the critical point, as this is the
 region which is usually hardest to reproduce using machine learning methods.
 The histograms and averages are demonstrated to agree with exact results within
 statistical errors even in the case
 where a diffusion model is used which is a relatively bad approximation to the exact diffusion process.
 The quality of the approximation by the neural net, and its simulation through
 a discretised solution of an SDE is monitored through the effective sample size (ESS).
 If the ESS decays too low, resamplings are used to rejuvenate the sample.

 We also investigate an improved update for the denoising process which is
 correct up to 2nd order in the stepsize. This allows for much larger stepsizes in the numerical integration of the denoising
 process, however, for large lattice models an approximation
  to the weight updates must be used, and an extrapolation to zero stepsizes becomes necessary.

  In summary, a new method is proposed for sampling using diffusion models.
  The method incurs relatively low additional costs to
  the simulations and ensures accurate results in spite of the nonzero stepsize
  solving the diffusion process and the inexact drift terms (``score'') that the neural network provides.
  The long term goal of this setup is therefore to enable sampling distributions where
  the action calculation is relatively cheap, but creating useful proposals in the
  Markov Chain Monte Carlo setup can be very costly (e.g. QCD).


\subsection*{Acknowledgments}

The numerical calculations were done on GSC, the HPC cluster at the University of Graz.

\footnotesize
\bibliographystyle{utphys}

\bibliography{../mybib}
  
\end{document}